# The effect of short-term exposure to the natural environment on depressive mood: A systematic review and meta-analysis

Hannah Roberts, Caspar van Lissa, Paulien Hagedoorn, Ian Kellar, Marco Helbich

**Abstract**

**Background**: Research suggests that exposure to the natural environment can improve mood, however, current reviews are limited in scope and there is little understanding of moderators.

**Objective**: To conduct a comprehensive systematic review and meta-analysis of the evidence for the effect of short-term exposure to the natural environment on depressive mood.

**Methods**: Five databases were systematically searched for relevant studies published up to March 2018. Risk of bias was evaluated using the Cochrane Risk of Bias (ROB) tool 1.0 and the Risk of Bias in Non-Randomised Studies of Interventions (ROBINS-I) tool where appropriate. The Grades of Recommendation, Assessment, Development, and Evaluation (GRADE) approach was used to assess the quality of evidence overall. A random-effects meta-analysis was performed. 20 potential moderators of the effect size were coded and the machine learning-based MetaForest algorithm was used to identify relevant moderators. These were then entered into a meta-regression.

**Results**: 33 studies met the inclusion criteria. Effect sizes ranged from −2.30 to 0.84, with a pooled effect size of $\gamma$ = -0.30 95% CI [-0.50 to -0.10]. However, there was significant residual heterogeneity between studies and risk of bias was high. Type of natural environment, type of built environment, gender mix of the sample, and region of study origin, among others, were identified as relevant moderators but were not significant when entered in a meta-regression. The quality of evidence was rated very low to low. An assessment of publication bias was inconclusive.

**Conclusions**: A small effect was found for reduction in depressive mood following exposure to the natural environment. However, the high risk of bias and low quality of studies limits confidence in the results. The variation in effect size also remains largely unexplained. It is recommended that future studies make use of reporting guidelines and aim to reduce the potential for bias where possible.

**Highlights**:

- A systematic review and meta-analysis is conducted of depressive mood following natural environment exposure
- Studies published up until March 2018 were searched; 33 studies met the eligibility criteria
- A small effect size for reduction in depressive mood following exposure is found
- Studies are found to be highly biased and of low or very low quality
- No significant moderators of the effect size are identified

# 1 Introduction

Depression is understood to have a lifetime prevalence of 10.8% among the global population (Lim et al., 2018). It is a leading contributor to the global disease burden (Ferrari et al., 2013), and at its worst can lead to suicide (Hawton, Casañas I Comabella, Haw, & Saunders, 2013). Characterised by depressed mood, loss of interest or enjoyment, and lack of energy, depression is estimated to affect over 300 million people worldwide (World Health Organization, 2017).

The natural environment is increasingly recognised as a potential buffer to poor mental health (Gascon et al., 2015; Houlden, Weich, de Albuquerque, Jarvis, & Rees, 2018; Rautio, Filatova, Lehtiniemi, & Miettunen, 2017; Van Den Berg et al., 2015). There are a number of existing theories linking green space and health (Hartig, Mitchell, de Vries, & Frumkin, 2014; Nieuwenhuijsen, Khreis, Triguero-Mas, Gascon, & Dadvand, 2017); two focus on the effects on mental health specifically: attention restoration theory (Kaplan, 1995) and stress reduction theory (Ulrich et al., 1991). Attention restoration theory proposes that the natural environment provides a 'soft fascination' whereby a person can pay attention without effort. Stress reduction theory suggests that the presence of nature brings about a psycho-evolutionary response related to safety and survival, and therefore produces positive emotions.

These pathways have received support in the literature: a number of cross-sectional studies (Beyer, Kaltenbach, Szabo, Bogar, & Nieto, 2014; Gascon et al., 2018; McEachan et al., 2015; Reklaitiene et al., 2014) and a few longitudinal studies (Alcock, White, Wheeler, Fleming, & Depledge, 2014; Astell-Burt, Mitchell, & Hartig, 2014) have found a relationship between increased surrounding green space and reduced risk of depression, and suicide also (Helbich et al., 2018; Min et al., 2017). Moreover, a recent systematic review of 28 studies found limited evidence of a causal relationship between quantity of and access to surrounding residential green space and mental health in adults (Gascon et al., 2015). However, it is unclear in these studies whether the association between green space and mental health is the result of use of green space or via another mechanism (Van den Bosch & Meyer-Lindenberg, 2019). Indeed, viewing green space from an indoor environment has been shown to have beneficial mental health effects (Ulrich, 1984; Ulrich et al., 1991). There is also a risk of self-selection bias, whereby associations might be attributed to those with fewer mental health problems moving into greener neighbourhoods.

Nevertheless, studies examining the relationship between direct exposure to the natural environment and mood have shown improvements after only a short period of time (Barton & Pretty, 2010; Shanahan et al., 2016; Shanahan, Fuller, Bush, Lin, & Gaston, 2015). Barton and Pretty (2010) assessed the effect on mood from exercising in nature and reported that the greatest benefit to mood came following 5 minutes of exercise, with diminishing returns thereafter. Moreover, Shanahan et al. (2016) found that the odds of reporting depression were significantly lower when respondents visited green space for an average of 30 minutes or more. In this way, short-term visits to the natural environment might represent a cheap and feasible intervention to improve mood. A clear understanding of the evidence base is then necessary in order to develop effective interventions.

Previous reviews of this topic are restricted in scope. For example, Lee et al. (2017) and Oh et al. (2017) examine the effect on depressive symptoms and health and well-being respectively, but both consider exposure to a specific type of natural environment only – forests. This disregards current research that considers the impact of other natural environments, such as parks (Song et al., 2013, 2014), agricultural land (Lee, Park, Ohira, Kagawa, & Miyazaki, 2015) and streetscape greenery (Helbich et al., 2019). Hansen, Jones, and Tocchini (2017) also review the effects of forest therapy on physical and mental health, but only include studies from Japan or China. While the concept of forest-bathing originated in Japan, resulting in a large proportion of this research coming from there, the number of studies from Europe and North America that also examine the effects of exposure to the natural environment is growing (Stigsdotter, Corazon, Sidenius, Kristiansen, & Grahn, 2017; Triguero-Mas et al., 2017). Lastly, McMahan and Estes (2015) include all types of natural environment and investigate its effect on positive and negative mood, but exclude all studies that are not randomised controlled studies. In sum, previous reviews are narrow in focus: an overall understanding of how short-term exposure to the natural environment affects depressive mood is missing in the literature.

In this review the findings of previous reviews are built on and the scope extended by including studies of varying designs and place of origin, and also adopting a wider understanding of 'natural environment'. All types of green space are considered – any open land with natural vegetation, planned or otherwise. A more complete picture of the effect of the natural environment on depressive mood is therefore presented. Blue space is not included in this review as a systematic review that considers blue space and mental health outcomes has recently been published (Gascon, Zijlema, Vert, White, & Nieuwenhuijsen, 2017).

Furthermore, a meta-analysis is conducted and an exploratory approach to moderator analysis is used. The machine learning-based MetaForest algorithm is applied to identify relevant moderators of the effect size (van Lissa, 2017). Moderators entered include age, gender mix of sample, present health condition, type of natural environment and region of study origin. This technique has not previously been applied in the green space-health literature, and therefore provides a novel contribution to a rapidly expanding field of research.

This systematic review and meta-analysis aims to summarise and critically appraise the evidence on the effect of short-term exposure to the natural environment on depressive mood. A secondary aim was to identify any potential moderators of this relationship. The final aim was to evaluate the quality of the evidence available.

## 2 Methods

The protocol was registered on PROSPERO (Roberts, Hagedoorn, Kellar, & Helbich, 2018). The review followed PRISMA guidelines (Moher, Liberati, Tetzlaff, Altman, & Group, 2009) (for checklist see Table S1 in the supplementary materials).

2.1 Eligibility criteria

Only original peer-reviewed research was eligible; abstracts, conference proceedings and grey literature were excluded. All geographical areas were eligible, but only references written in English were included.

2.2 PECO statement

A PECO was developed in order to inform the development of search strategies and guide the screening of relevant studies.

Population: Any human adult population

Exposure: Exposure to the natural environment

Comparator: Exposure to the built environment

Outcome: Depressive mood

In terms of populations considered, any adult population was eligible, regardless of physical or mental health status. Exposure was defined by placement of participants in direct contact with the exposure environment, within the context of a (randomised or non-randomised) trial (e.g. crossover, parallel group, factorial). Exposure duration was not limited, but based on a previous systematic review of the effects of short-term, direct exposure to the natural environment on health and well-being, it was expected that a single exposure would last approximately one hour with exceptions for those that applied repeated exposures (Bowler, Buyung-Ali, Knight, & Pullin, 2010). Representations of an environment using virtual reality, pictures or video were excluded. Environments were deemed as 'natural' if they were defined by a high level of greenery and had not been extensively modified by human activity. In contrast, the built environment was defined as a predominantly man-made environment with a low level of greenery. Studies where participants were exposed to more than two environments but included the natural and built environment were eligible, however, only data from the natural and built environment were included in the meta-analysis.

The primary outcome was depressive mood. This should be measured quantitatively either by the researcher(s) or the participant. The measure must refer to the current emotional state, rather than depressive mood over a longer preceding period. It might be measured independently, or as part of a wider mood or affect assessment. Measurements could be recorded pre- and post-exposure, or post-exposure only. Studies that measured well-being or quality of life were excluded as they were understood to be concepts distinct from depression.

2.3 Search strategy

A literature search was conducted on five databases: Medline, PsychINFO, Embase, Scopus and Web of Science from inception to March 2018. Search terms were related to the natural environment (such as 'natural environment', 'green space', 'open space' or 'park') and to depression or depressive mood (such as 'depression', 'depressive symptoms', 'mood disorder', and 'mental health'). The full search strategies are available in the supplementary material.

2.4 Study selection

Records from each database were downloaded and merged in Endnote. Duplicates were removed. The titles and where available, abstracts, of the remaining references were screened independently by two researchers according to the PECO statement and eligibility criteria. Percent agreement was 99%. All eligible references were then evaluated at the full-text level. Full papers were screened independently by two authors using the same criteria. Percent agreement was 98%. Reference lists of included studies were also hand-searched for relevant studies.

2.5 Data extraction

Relevant data were extracted by the first author using a standardised form. This included information on the study design, participant information, intervention and control environments, procedural descriptions, outcomes and outcome measures, results and conclusions. All data extraction forms were checked by a second researcher, and any disagreement was resolved through discussion until consensus was reached.

2.6 Quality assessment

The Cochrane Risk of Bias (ROB) Tool 1.0 (Higgins et al., 2011) was used to assess included randomised studies. The tool gives an overall risk of bias for randomised trials by scoring them across seven domains: random sequence generation, allocation concealment, blinding of participants and personnel, blinding of outcome assessment, incomplete outcome data, selective reporting, and any other sources of bias. The Risk of Bias In Non-Randomised Studies of Interventions (ROBINS-I) (Sterne et al., 2016) tool was used for non-randomised studies. Studies are scored on seven domains: confounding, participant selection, classification of interventions, deviation from intended intervention, missing data, measurement of outcomes, and reporting bias. All included studies were independently evaluated by two researchers. Appraisals were discussed between the researchers until consensus was reached.

Quality of evidence was assessed using the Grading of Recommendations Assessment, Development, and Evaluation (GRADE) guidelines (GRADE Working Group, 2004; Guyatt et al., 2008). These guidelines apply a set of predetermined domains that either increase or decrease the level of confidence in the evidence. Domains that reduce confidence in the evidence are: risk of bias, inconsistency of results, indirectness of evidence, imprecision, and publication bias. On the other hand, a large magnitude of effect, confounding that increases effect magnitude, and a dose-response gradient can increase confidence. Two researchers discussed the domains for each outcome until consensus was reached.

2.7 Data synthesis

*Narrative summary*

Studies are first narratively discussed in the context of the type of engagement performed by participants in the exposure environment: active engagement (e.g. walk, run), passive engagement (e.g. sit, stand), or a combination of both.

*Meta-analysis*

All studies were considered for inclusion in a meta-analysis. Two types of effect size were calculated, where appropriate: standardized mean difference (SMD) (Hedges' G) for randomized controlled experiments (Hedges, 1981), and standardized mean change rates (SMCR) for pretest- posttest designs (Morris & DeShon, 2002). Hedges' G is recommended when sample sizes are below 20, and expresses the difference of the means in units of the pooled standard deviation. Furthermore, it can be interpreted similarly to Cohen's d (e.g. 0.2 refers to a 'small' effect) (Fritz, Morris, & Richler, 2012). Mean (or mean change pre- and post-exposure), standard deviation (or standard error) and cell count (n) for all depression outcomes in each included study were extracted. In the first instance, data was extracted directly from the studies. When the data was not available, authors were contacted for further information. Where data could not be provided or contact could not be made, data was extracted from figures using an online ruler (A Ruler for Windows). Two studies did not report the standard deviation (Hartig, Mang, & Evans, 1991; Hartig, Nyberg, Nilsson, & Garling, 1999), therefore an estimate was derived by taking the average from similar papers that used the same outcome measure. One study released a corrigendum after the literature search had been conducted, therefore the author was contacted and the correct data used in the meta-analysis (Lee et al., 2019). It was not possible to retrieve data for two studies (Li et al., 2016; Perkins, Searight, & Ratwik, 2011).

Only data from the natural environment and built environment were considered. Data from other settings e.g. blue space, passive lab setting were not included in meta-analysis. Two studies had multiple 'green' environments: Sonntag-Öström et al. (2014) examined differences across a spruce forest, a forest with a lake, and a forest with a rocky outcrop, and Tyrväinen et al. (2014) considered both a park and a forest. The first environment listed was selected for both studies. Moreover, it was felt that the lake and 'rocky outcrop' present in the alternative environments of Sonntag-Öström et al. (2014) may act as confounders. Only data from timepoints closest to the start and end time of the exposure were extracted; baseline or follow-up measures, or measurements taken during exposure, were not included in analysis. For cases of multiple exposure to the same environment, data was extracted from immediately before the first exposure, and immediately after the final exposure.

*Moderator analysis*

For each study a number of potential moderators were coded, for example study design, region of study origin, and mean age of the sample. For the full list of moderators, see Table 1. For studies that had multiple exposures to the same environment, exposure time was measured cumulatively.

**Table 1.** Moderators coded for meta-analysis

| Moderator | Potential codes |
|---|---|
| Study design | Crossover design |
| | Parallel groups |
| | Factorial design |
| Region of study origin | Asia |
| | Europe |
| | US |
| Mean age of sample | - |
| Student sample | Yes |
| | No |
| Gender mix of sample | Male |
| | Female |
| | Mixed |
| Female (%) | - |
| Health condition of sample | Healthy |
| | Poor mental health |
| | High blood pressure |
| | Chronic heart failure |
| | Chronic obstructive pulmonary disease |
| Total exposure time (minutes) | - |
| Time between crossover to other environment (if appropriate) | Same day |
| | Next day |
| | Next week |
| | Longer |
| Type of natural environment | Forest |
| | Park |
| | Biodiverse area |
| | Agricultural area |
| Type of built environment | Downtown |
| | Residential |
| | Other |
| Number of natural environments | - |
| Number of built environments | - |
| Baseline measurement | Yes |
| | No |
| Measurement conducted at environment | Post-exposure only |
| | Pre and post-exposure (either side multiple exposures) |
| | Pre and post-exposure for each exposure |
| Measurement conducted during exposure | Yes |
| | No |
| Follow-up measurement | Yes |
| | No |
| Activity category | Passive |
| | Active |
| | Mixed |
| Primary depression measure | - |
| Secondary depression measure (if appropriate) | - |

Missingness was very limited; three variables (gender mix, proportion of female, time between environments) had some missing values, ranging from 5-32%. Since complete data was required for analysis, single imputation was applied using a non-parametric missing value imputation by means of a random forest algorithm (Stekhoven & Bühlmann, 2012).

The number of moderators coded (*n*=20) was large relative to the sample size. Consequently, including all moderators in a meta-regression risks overfitting the model. Therefore we apply the R package 'metaforest' (van Lissa, 2017); an exploratory approach to identify potentially relevant moderators in meta-analysis. The approach is based on the machine-learning algorithm 'random forests', which are robust to overfitting. First, the approach ranks moderators in terms of their importance in predicting the effect size. Second, partial dependence plots are produced which visualize the association of each moderator with the effect size, while accounting for the average effect of all other moderators. Lastly, a measure of 'predictive performance', or the 'out-of-bag' $R^2_{oob}$, is calculated for each moderator. In other words, an estimate is given of how much variance the moderators would explain if a new sample of data were provided. Moderators that consistently displayed negative variable importance (i.e., that showed a reduction in predictive performance) were dropped. Moderators that improved predictive performance were then entered into a linear meta-regression in order to understand their association with the effect size. For categorical variables, contrast coding is applied, such that the levels of one variable are compared with the mean of the subsequent levels. For ordinal variables, orthogonal polynomial coding is applied, and the linear, quadratic and cubic trends considered.

Publication bias was first assessed by visual examination of funnel plots. Standard error was used as the measure of study size, plotted on the vertical axis, with effect estimates plotted on the horizontal axis (Sterne & Egger, 2001). A symmetrical, inverted funnel indicates absence of bias. In addition, funnel plot asymmetry was tested using Begg's test, which examines the association between the effect estimates and their variances. Lastly, file drawer analysis was completed (Rosenthal, 1979). This calculates the number of studies averaging null results that would have to be added to nullify the summary effect (i.e. reduce the combined significance level (p-value) to a target alpha level (e.g. 0.05)).

The meta-analysis was completed in R (R Core Team, 2018) using the metafor (Viechtbauer, 2010) and MetaForest packages (van Lissa, 2017). The full reproducible code is available online at: OSF.IO/S2JV4.

## 3 Results

Key characteristics of each included study are shown in Table 2. The initial database search yielded 8,958 results, of which 2,336 were removed as duplicates. 6,622 titles and abstracts were screened and those deemed potentially relevant were retrieved as full texts. 76 studies were identified for full text screening. A further nine were retrieved from checking reference lists. In total, 33 studies met the inclusion criteria. For the flow diagram of this process, see Figure S1.

Study Characteristics

Studies came from 10 different countries, with most originating from Japan ($n$=13) or the US ($n$=5). Nine were published in Europe. The majority were randomised crossover studies ($n$=16) or non-randomised crossover studies ($n$=5). Seven studies used parallel groups, three had a factorial design and two were single-group crossover studies.

**Table 2.** Study characteristics

| Reference | Country | Sample | Design | Intervention | Setting(s) | Depression Measure(s) | Outcome |
|---|---|---|---|---|---|---|---|
| **Active engagement during exposure** | | | | | | | |
| Berman et al., 2013 | USA | *n*=20 (8 men) Mean age: 26 MDD diagnosis | Randomised crossover | 2.8 mile, 50-55 min walk. | Natural environment: Ann Arbor Arboretum Built environment: downtown Ann Arbor. | PANAS | - Decreases in negative affect observed after both the nature walk and the urban walk. <br> - No significant effect of location on negative affect, but significant main effect of time. |
| Bodin & Hartig, 2003 | Sweden | *n*=12 (6 men) Male mean age: 39.7, SD 6.1 Female mean age: 37.0, SD 7.0 | Randomised crossover (4-period 2-treatment) | 60 min running route (max 14km). | Natural environment: nature reserve. Built environment: route through Uppsala city | NMS (anxiety/depression, anger subscales) | - The declines in anxiety/depression and anger subscales from pre-test to post-run were significant. <br> - Change in NMS in the park environment was not significantly different to change in the urban environment. |
| Gidlow et al., 2016 | England (UK) | *n*=38 (23 men) Mean age: 40.9, SD 17.6. Healthy | Randomised crossover (3-period, 3-treatment) | 30 min walk. | Natural environment: country park within the city Built environment: quiet residential streets Blue site: footpath besides a canal | BRUMS | - Mood improved from baseline in all environments, with a significant main effect of time on TMD, post-walk and 30 minutes after leaving the environment. <br> - No significant main effect for environment and no significant environment*time interaction effect. |
| Hartig et al., 1991 (Study 2) | USA | *n*=34 (17 men) Mean age: 20 College students | Randomised parallel group | 40 min walk (sitting in passive site). | Natural environment: Santiago Oaks Regional Park, California Built environment: Santa Ana, California Passive site: University of California campus | OHS; ZIPERS | - The mean sadness score on ZIPERS was not significantly different for the natural environment group compared to the other groups. |
| Jia et al., 2016 | China | *n*=18 COPD patients Mean age: 70.1 | Randomised parallel group | 90 min walk Repeated in the morning and afternoon for 3 days. | Natural environment: White Horse Mountain National Forest Park. Built environment: Hangzhou city | 65-item POMS | - Depression was significantly reduced in the forest group between pre and post-exposure.. <br> - No significant difference seen in the urban group, and score not significantly different between groups. |
| Johansson, Hartig, & Staats, 2011 | Sweden | *n*=20 Mean age: 23.3 Students | 2x2 factorial design | 40 min walk With/without friend as within-subject factor. | Natural environment: Municipal park | NMS | - Statistically significant main effect of time for all affect measures. <br> - Negative affect was not significantly modified by environment or social context. |

| | | | | | Built environment: Street walk in mixed land use area | | |
|---|---|---|---|---|---|---|---|
| Lee et al., 2014 | Japan | *n*=48 men<br>Mean age: 21.1<br>Mean BMI: 21.3 | Randomised crossover | 12-15 min walk. | Natural environment: forest<br>Built environment: urbanised area near forest | 30-item POMS | - No significant change before and after forest and urban walks on the POMS depression subscale. |
| Li et al., 2016 | Japan | *n*=19 men<br>Mean age: 51.2, SD 8.8<br>High-normal or hypertension | Single group crossover | 2.6km, 80 min walk. Repeated in morning and afternoon. Visited other site one week later. | Natural environment: forest park<br>Built environment: urban area in Nagano prefecture. | 65-item POMS | - Significant decrease in the depression subscale after walking in the forest in the morning compared to before walking.<br>- No information on afternoon or Built environment; assumed results were not significant. |
| Mao, Cao, et al., 2012 | China | *n*=34<br>Age range for inclusion: 60-75 years.<br>Patients with essential hypertension. | Randomised parallel group | 90min walk with a 20 min break. Repeated in the morning and afternoon for 7 days. | Natural environment: White Horse Mountain National Forest Park.<br>Built environment: location in Hangzhou city | 65-item POMS | - The forest group had a significantly lower depression subscale score between baseline and post-intervention.<br>- No significant change in the urban group. |
| Mao, Lan, et al., 2012 | China | *n*=20 men<br>Mean age: 20.79, SD 0.54.<br>Students | Randomised parallel group | 90 min walk with a 10 min break. Repeated in the afternoon. | Natural environment: evergreen forest in Hangzhou, China<br>Built environment: nearby urban area | 65-item POMS | - Depression subscale score was significantly lower than that of the urban group post-intervention. |
| Mao et al., 2017 | China | *n*=33<br>Mean age: 72.2<br>CHF patients | Randomised parallel group | 90 min walk, twice a day for four days. | Natural environment: forest site in Pan'an county<br>Built environment: downtown area of Hangzhou | 65-item POMS | - Significant decrease for the forest group in the depression subscale compared with baseline score.<br>- Post-intervention score for this subscale also significantly lower than the post-intervention score for the urban group. |
| Perkins et al., 2011 | USA | *n*=26 (7 men)<br>Age range 19-24<br>Students | Parallel group<br>(3 groups) | 20 min walk | Natural environment: wooded trail<br>Built environments: mixed residential/business neighbourhood; parking lot. | 65-item POMS | - For all settings, change in depression score pre and post-intervention not significantly different. |

| Roe & Aspinall, 2011 (Study 2) | Scotland (UK) | n=24 (11 in good mental health, 13 in poor mental health) Poor health: clinically diagnosed mental health problem | Non-randomised crossover | 60 min walk. Repeated in other setting one week later. | Natural environment: Plean Country Park, Stirlingshire Built environment: Stirling town centre | MACL (hedonic tone, energy, stress). | - Significant positive change in mood following the rural walk; no significant change following the urban walk for the good health group.<br>- Significant positive change in mood for both the urban and rural walks for the poor health group. |
|---|---|---|---|---|---|---|---|
| Shin, Shin, Yeoun, & Kim, 2011 | Korea | n=60 (35 men, 25 women) Mean age 23.27 Students | Randomised crossover | 4.5km, 50-55 min walk. Repeated in other setting one week later. | Natural environment: Natural forested park Built environment: downtown Cheongj | 65-item POMS | - All POMS subscales, including depression, significantly improved following exposure to the forest.<br>- Depression subscale score increased following the urban exposure, but this was not significant. |
| Song et al., 2013 | Japan | n=13 men Mean age 22.5, SD 3.1 Students | Non-randomised crossover | 15 min walk. Rested for 20 mins then repeated in other setting. | Natural environment: Kashiwanoha Park in Chiba, Japan Built environment: city area around the park | 30-item POMS | - No significant difference between settings was observed in the POMS depression subscale. |
| Song et al., 2014 | Japan | n=17 men Mean age: 1.2, SD 1.7 Students | Non-randomised crossover | 15 min walk. Repeated in other setting one hour later. | Natural environment: Kashiwanoha park in Kashiwa City Built environment: city area around the park | 30-item POMS | - No significant difference between settings was observed in the POMS depression subscale. |
| Song, Ikei, Kobayashi, et al., 2015 | Japan | n=19 men Mean age: 58, SD 10.6 High-normal or hypertensive blood pressure | Randomised crossover | 17 min walk. Repeated in other setting the next day. | Natural environment: Akasawa natural recreation forest Built environment: Ina city | 30-item POMS | - Depression subscale of POMS was significantly lower after walking in forest than walking in the urban area. |
| Song, Ikei, Igarashi, Takagaki, & Miyazaki, 2015 | Japan | n=23 men Mean age 22.3, SD 1.2 Students | Non-randomised crossover | 15 min walk. Rested for 20 min then repeated in other setting. | Natural environment: Kashiwa-no-ha Park Built environment: City area around the park | 30-item POMS | - No significant difference between settings was observed in the POMS depression subscale. |
| Stigsdotter, Corazon, Sidenius, Kristiansen, & Grahn, 2017 | Denmark | n=51 women Age range 20-36 University students | Non-randomised crossover | 15 min walk. Other setting visited within 2 weeks. | Natural environment: Octavia health forest Built environment: historic downtown area of Copenhagen | 65-item POMS | - Depression scores reduced after both the forest and urban walk, however the change was significant only for the urban walk.<br>- The depression score was significantly higher in the urban environment pre-exposure compared to the forest. |

| Triguero-Mas, Gidlow, et al., 2017 | Catalonia (Spain) | n=26 Eligible if Mental Health Inventory score in the lower 50th percentile | Randomised case-crossover | Participants asked to "spend time" in environment. 30 min + 180 min. | Natural environment: Collserola Natural Park Built environment: Eixample neighbourhood in Barcelona Blue site: Castelldefels beach | 29-item POMS | - TMD was significantly lower in the green site compared with the built environment. - TMD was significantly lower in the blue site compared to the green site. |
|---|---|---|---|---|---|---|---|
| **Passive engagement during exposure** | | | | | | | |
| Bielinis, Takayama, Boiko, Omelan, & Bielinis, 2017 | Poland | n=62 (36 men) Mean age: 21.45, SD 0.18 Students | Randomised parallel group | 15 min standing in environment. Repeated in other setting in the afternoon. | Natural environment: urban forest (deciduous, broad-leaved) Built environment: Olsztyn city, Poland | PANAS, 65-item POMS | - Depression score on POMS after exposure to forest environment significantly lower to score after exposure to the urban area. - Negative affect was significantly higher following exposure to the urban area compared to after exposure to the forest environment. |
| Hartig et al., 1999 (Study 3) | USA | n=101 Mean age: 20.6 Students | 2x2 factorial design | Participants had 5 mins to draw environment, then 10 mins sitting. | Natural environment: Botanical gardens on Berkeley campus Urban: Busy traffic intersection | PANAS; ZIPERS | - Those in the natural environment reported lower sadness in ZIPERS and negative affect in PANAS than those in the urban environment. - The difference between groups was not significant. |
| Igarashi et al., 2015 | Japan | n=17 women Mean age: 46.1 Mean BMI: 21.4 | Randomised crossover | 10 mins sitting. Repeated in other environment immediately after first | Natural environment: kiwifruit orchard Built environment: building site | 30-item POMS | - Score on depression subscale on POMS significantly lower following orchard visit compared to the building site visit. |
| Joung et al., 2015 | Korea | n=8 Mean age: 22.0. Students | Single group crossover | 15 mins sitting | Natural environment: local forest Built environment: Daejeon city (conducted on rooftop) | POMS | - POMS depression subscale was not significantly different between environments after exposure. |
| Lee et al., 2011 | Japan | n=12 men Mean age: 21 Mean BMI: 22.5 Students | Randomised crossover | 15 mins sitting Repeated in other setting the next day | Natural environment: forest in Tsurui village, Hokkaido Built environment: commercial area of Kushiro town, Hokkaido | 30-item POMS | - Depression subscale score was higher in the urban area than the forest following exposure. No significant differences were observed. |

| Lee et al., 2015 | Japan | n=12 men<br>Mean age: 22.3<br>Students | Randomised crossover | 15 mins sitting in environment<br>Repeated in other setting the next day | Natural environment: paddy field in Ukiha city in southern Japan<br>Built environment: Hakata railway station | POMS | - Depression subscale score was significantly lower for the rural environment compared to the urban environment post-exposure. |
|---|---|---|---|---|---|---|---|
| Sonntag-Öström et al., 2014 | Sweden | n=20 women<br>Mean age: 41.6, SD 7.3<br>Mean level of burnout: 5.7 (7-point Burnout Questionnaire) | Randomised crossover (4-period, 4-treatment) | 40 mins sitting in environment | Multiple natural environments: Forest by lake, rocky outcrop, spruce forest<br>Built environment: Umea city | Adapted mood questionnaire based on POMS and ZIPERS | - Significant differences between different environments found for all mood scales, except exhausted-alert.<br>- Participants rated higher on the scales relaxed, happy, harmonious, peaceful and clearheaded in all forest environments compared to the city. |
| Tsunetsugu et al., 2013 | Japan | n=46 males<br>Mean age: 21.1, SD 1.1<br>Students | Non-randomised crossover | 15 mins sitting in environment<br>Repeated next day in other environment. | Natural environment: four forests in central and Western Japan<br>Built environment: four urban areas in central and Western Japan | 65-item POMS | - No significant change in the depression subscale score of POMS. |
| **Combination of active and passive engagement during exposure** | | | | | | | |
| Hartig, Evans, Jamner, Davis, & Gärling, 2003 | USA | n=112 (56 men) Students<br>Mean age 20.8 SD 3.7<br>Healthy | 2x2 factorial design | Cognitive task prior to walk as within-subject factor.<br>10 mins sitting, then 50 mins walk. | Natural environment: Audubon Society's Starr Ranch Sanctuary, California<br>Built environment: Orange city, California | OHS; ZIPERS | - Those that walked in the nature reserve experienced more positive emotion than those walking in the urban environment, within the no-task condition. The main effect of environment among the task subjects was not significant. |
| Park, Tsunetsugu, Kasetani, Kagawa, & Miyazaki, 2010 | Japan | n=280<br>Mean age: 21.7, SD 1.5<br>Students | Randomised crossover | 14 mins sitting, 16 mins walk.<br>Repeated in other environment the next day. | 12 forest sites and 12 urban areas across Japan | POMS | - The POMS depression subscale score significantly decreased following the viewing of the city area and improved in the forest area.<br>- When walking, the change in the average POMS depression subscale score was also significantly different between the forest and city areas. |
| Park et al., 2011 | Japan | n=168 men<br>Mean age: 20.4, SD 4.1<br>Students | Randomised crossover | 15 mins sitting in the morning, 15 mins walk in the afternoon.<br>Repeated in other environment the next day | 14 forest sites and 14 urban areas across Japan | POMS | - After performing both activities, TMD scores were significantly lower for the forest areas than for the urban areas. However, no significant differences were observed for the depression subscale. |

| Takayama et al., 2014 | Japan | *n*=45 men | Randomised crossover | 15 mins walk in the morning, 15 mins viewing in the afternoon. Repeated in other environment the next day | 4 forest sites and 4 urban areas across Japan. | PANAS; POMS | - Depression score was significantly lower after the experiment in the forest environment; no significant change to depression in the urban environment.<br>- No significant difference between environments for the depression subscale post-exposure<br>- No statistical difference between before the experiment and after viewing either in negative affect or positive affect. |
|---|---|---|---|---|---|---|---|
| Tyrväinen et al., 2014 | Finland | *n*=77 (6 men) Mean age: 47.6 | Randomised crossover | 15 minutes viewing, 30 mins walk (approx. 2km) At least one week between each visit | Multiple green sites: Alppipuisto (urban park), and Keskuspiusto (large urban woodland) Built environment: Helsinki city centre | PANAS | - People had fewer negative emotions in the forest compared to the city<br>- Interaction between environment and time was not significant. |

*Note.* MACL: Mood Adjective Checklist; NMS: Negative Mood Scale; OHS: Overall Happiness Scale; PANAS: Positive and Negative Affect Schedule; POMS: Profile of Mood States; TMD: Total Mood Disturbance, ZIPERS: Zuckerman Inventory of Personal Reactions

*Participants*

Sample sizes ranged from 8 participants (Joung et al., 2015) to 280 (Park et al., 2010). On the whole samples were small with 76% of studies (*n*=25) including less than 50 participants. Participants were typically young, with just over half of studies (*n*=18) recruiting college or university students. Some studies specified a clinical population. This included persons with: major depressive disorder (Berman et al., 2013), high-normal or hypertension (Li et al., 2016; Mao et al., 2012a; Song et al., 2015), congestive heart failure (CHF) (Mao et al., 2017), chronic obstructive pulmonary disease (COPD) (Jia et al., 2016), a mental health problem (Roe & Aspinall, 2011), a high level of burnout (Sonntag-Ostrom et al., 2014), and a poor Mental Health Inventory score (Triguero-Mas et al., 2017).

*Intervention*

Exposure time ranged from 10 minutes to 90 minutes, with 15 minutes being the most common (*n*=11). Some studies had multiple exposures within one day (Li et al., 2016; Park et al., 2011; Takayama et al., 2014; Triguero-Mas et al., 2017). Three studies from the same researcher had a considerably longer exposure time whereby participants completed a walk in the morning and afternoon for a period of 2 days (Mao et al., 2012b), 4 days (Mao et al., 2017) and 7 days (Mao et al., 2012a).

For crossover studies, most often the second environment would be visited the following day (*n*=8), Nine studies specified a length of time ranging from "at least five days apart" to "within two weeks". One study only indicated that visits were undertaken within the same season (Sonntag-Ostrom et al., 2014). In contrast, in five studies participants visited both environments on the same day, with one study giving participants only one minute to turnaround to face the other environment and three minutes to rest before measurements began again (Igarashi et al., 2015).

Sixteen studies had participants actively engage with the environment – most asked participants to walk, and one asked participants to complete a run (Bodin and Hartig, 2003). One study allowed participants to choose what to do, only asking them to "spend time" in the environment (Triguero-Mas et al., 2017). This was coded as active engagement as it was assumed participants would move somewhat within the exposure area. A second group of studies had participants passively engage: seven asked participants to sit and view the environment, and one had participants stand due to the cold weather (Bielenis et al., 2018). Five used a combined approach whereby participants walked in and then viewed the environment or vice versa (Hartig et al., 2003; Park et al., 2010; Park et al., 2011; Takayama et al., 2014; Tyrvainen et al., 2014).

*Outcome Measures*

The most frequently used mood measure was the Profile of Mood States (POMS) (*n*=22). Also used was the Positive and Negative Affect Scale (PANAS) (*n*=5) and Zuckerman Inventory of Personal

Reactions (ZIPERS) (*n*=3). The Overall Happiness Scale (OHS) and Negative Mood Scale (NMS) were used twice each, and the Mood Adjective Checklist (MACL) and Brunel Mood Scale (BRUMS) were all used once. One study used a bespoke questionnaire that was based on the POMS and ZIPERS (Sonntag-Ostrom et al., 2014).

Most studies (*n*=24) took mood measurements pre- and post-exposure (including studies that had more than one exposure per environment). Nine studies measured mood at post-exposure only. In addition to pre-post measurement, five studies took a baseline measurement before traveling to the exposure environment (Gidlow et al., 2016; Lee et al., 2011, Lee et al., 2015; Park et al., 2010; Triguero-Mas et al., 2017). Furthermore, two studies took a second post-exposure measurement: Gidlow et al. (2016) measured mood immediately after the exposure had finished, and then again 30 minutes later, while Triguero-Mas et al. (2017) completed a final measurement upon return to the lab. Two studies captured mood during the exposure period (Hartig et al. 2003; Li et al., 2016).

*Setting*

Most studies used forests as their natural environment (*n*=16), followed by urban or country parks (*n*=11). Four used natural environments characterised by their biodiversity (nature reserve, botanical garden) and two used more agricultural settings (kiwifruit orchard: Igarashi et al., 2015, paddy field: Lee et al., 2015).

For the comparative built environment, most studies described a location in a downtown, urban area (*n*=27). One study indicated the location was in an urban area, but participants viewed the area from a rooftop (Joung et al. 2015). Two studies used a residential street (Gidlow et al., 2016; Triguero-Mas et al., 2017), one used a building site (Igarashi et al., 2015), and one a railway station (Lee et al., 2015). Some studies had additional environments that were not explored in this review: a canal path (Gidlow et al., 2016), a beach (Triguero-Mas et al., 2017), forest with rocky outcrop, forest by a lake (Sonntag-Ostrom et al., 2014) and a lab setting (Hartig et al., 1991).

Risk of Bias

25 randomised studies were evaluated using the Cochrane Risk of Bias 1.0 tool (Higgins et al., 2011); 8 non-randomised studies were evaluated using the ROBINS-I (Sterne et al., 2016) (see Figure S2 and Table S2 respectively).

Concerning the randomised studies, two studies described their method of randomisation and therefore were assigned a low risk of bias in this domain (Gidlow et al., 2016; Sonntag-Ostrom et al., 2014). One study was given high risk of bias because participants were assigned to an exposure group based on participant availability (Triguero-Mas et al., 2017). Remaining studies did not describe their method of randomisation, and so were rated as unclear.

A number of important confounders were identified in the non-randomised studies, meaning six (of 8 total) were marked as having moderate or serious risk of bias. Many confounders were possible, but in completing the assessment particular attention was paid to: the weather, food, alcohol and caffeine consumption; social interaction with other participants or researchers; the environment participants were exposed to immediately before measurements started; and the length of time between the experimental and control environment exposures (if applicable). For example, Li et al. (2016) prohibited alcohol, caffeine and smoking during the study period, and participants were not allowed to speak to each other during their walk in the exposure environment. However, they state that the weather was sunny for the built environment exposure, and rainy and cloudy for the natural environment exposure. It was also not clear how the participants travelled to the exposure environments, therefore the type of environment they were exposed to prior to measurement and possible social interactions were not known. For these reasons, the study was marked as having serious risk of bias. Two studies gave too little information on the confounders listed to make an informed decision and were marked as 'no information' (Joung et al., 2015; Tsunetsugu et al., 2013).

All non-randomised studies were judged as low risk for selection bias because the selection of participants was not related to the intervention or the outcome. Due to the nature of the interventions it was judged that there was also little risk of misclassification of intervention and control sites. No study was deemed to have 'deviated from intended intervention'.

In terms of blinding, all studies were judged as highly biased: blinding is impossible due to the nature of the studies. Some studies attempted to minimise bias by not informing participants which environment would be visited first (Gidlow et al., 2016; Sonntag-Öström et al., 2014). Nevertheless, the outcome was subjective, and participants were likely to be aware of the hypothesis being tested.

It was judged likely that attrition was related to the outcome unless otherwise stated. For this reason, 12 of the randomised studies had a high risk of attrition bias. Studies that did not explain different numbers of participants reported in the methods and results were assigned an unclear rating ($n$=5). The remaining randomised studies ($n$=8) reported no drop outs, and therefore had a low risk of bias. Within the non-randomised studies, Roe and Aspinall (2011) noted that attrition was concentrated in their 'poor health' group only, therefore was rated as having serious risk of bias. All other studies had complete data or the proportion missing was limited.

Two non-randomised studies did not report full results, resulting in a serious risk of reporting bias. All other studies received an unclear (using ROB) or moderate (using ROBINS-I) risk of bias since full data was reported but did not have associated study protocols.

Narrative data synthesis

*Active engagement interventions (n=20)*

Eleven studies reported a significant decrease in depression pre and post-exposure to the natural environment. For example, Mao et al. (2012a) and (2017) had participants walk 90 minutes twice a day for 7 and 4 days respectively, both reporting a significant decrease in depressive mood in the forest environment compared to the pre-exposure score. Shin et al. (2011) had participants walk in a forest for 50-55 minutes, and in a city the following week. All POMS subscales, including depression, were found to significantly improve following the forest exposure. However, six of the eleven studies were not able to demonstrate that the change in mood was significantly different to that observed in the built environment (Berman et al. 2013; Bodin and Hartig, 2003; Gidlow et al., 2016; Jia et al., 2016; Johansson et al., 2011; Roe and Aspinall, 2011).

Four studies showed no significant change in mood pre and post exposure to a natural environment (Stigsdotter et al. 2017; Hartig et al., 1991; Lee et al., 2014; Perkins et al., 2011). Li et al. (2016) found a significant decrease in the depression subscale after walking in the forest compared to baseline, but no results from the built environment are presented.

Five studies assessed mood at post-exposure only (Mao et al., 2012b; Song et al., 2013; Song et al., 2014; Song et al., 2015a; Song et al., 2012b). Two reported that the POMS depression subscale score was significantly lower following the forest visit than the built environment visit (Mao et al., 2012b; Song et al., 2015a); the remaining three studies found no significant difference between environments.

*Passive engagement interventions (n=8)*

Four studies compared depressive mood pre- and post-exposure, with two finding a significant reduction. Bielinis et al. (2017) examined change in mood following winter forest bathing in young students. They were asked to stand in a forest and built environment for 15 minutes. The depression score was significantly lower after exposure to the forest compared to the built environment. Lee et al. (2015) also had an exposure period of 15 minutes. They report that depression was significantly lower in the rural environment post-exposure. There was no significant change in mood in the other two studies (Lee et al., 2011; Tsunetsugu et al., 2013).

Four studies measured post-exposure only, also with two reporting significant results. Igarashi et al. (2015) asked women to sit for ten minutes in an orchard, and then in a building site. Depression was rated significantly lower after sitting in the orchard than the building site. Sonntag-Ostrom et al. (2014) sent 20 female patients with exhaustion disorder to four different environments in Sweden, testing for mood, attention and physiological response. Environments were three forest environments (spruce forest, forest with a lake, forest with rocky outcrop) and a built environment. Patients reported feeling significantly more happy, relaxed, harmonious, peaceful and clearheaded in all forest environments compared to the city environment.

The remaining two studies (Hartig et al., 1999; Joung et al., 2015) did not find a significant difference in mood following exposure. Joung et al. (2015) note this may be explained by the fact that participants sat on a rooftop to observe an urban area, therefore preventing full immersion of the participant in the environment.

*Combination of active and passive engagement interventions (n=5)*

Park et al. (2010) and Takayama et al. (2014) both reported a significant reduction in the POMS depression subscale after a 15 minute walk and 15 minute viewing session in a forest. Park et al. (2011) followed the same procedure of 15 minutes walking and viewing in 14 forest and built environment sites across Japan, but no significant change in depressive mood was observed. Tyrvainen et al. (2014) compared results across three environments: an urban park, urban woodland, and city centre. People experienced fewer negative emotions in the woodland compared to the park and city centre, but there was no interaction between place and time. Similar results were found in Johansson et al. (2011), who compared participants walking in a park and down a street, and with and without a friend accompanying them. There was a significant main effect of time, but change in negative affect was not modified by environment or social context. Finally, Hartig et al. (2003) asked half of the participants to complete a cognitively demanding task. Those who did not complete the task experienced more positive emotion following the natural environment walk than the built environment walk, however for those who completed the task, there was no significant main effect of environment.

Meta-analysis

*Descriptive statistics*

Observed effect sizes ranged from -2.30 to 0.84. The unweighted mean effect size was $M_g = -0.29$, $SD = 0.60$, which can be interpreted as a small effect. Six studies reported two effect sizes (i.e. two outcomes). Because of this, a three-level meta-analysis was first used to estimate the amount of within-study- and between-studies variance (Van den Noortgate, López-López, Marín-Martínez, & Sánchez-Meca, 2014).

As indicated in Table 3, the within-study variance component did not differ significantly from zero, $\sigma_w^2 < 0.01$, 95% CI [0, 0.28]. The between-studies variance component, on the other hand, was significant, $\sigma_b^2 = 0.31$, 95% CI [0.09, 0.55]. Thus, the variation in observed effect sizes was primarily due to differences between studies. As the within-study variance component was near-zero, and the Akaike information criterion (AIC) and Bayesian Information Criterion (BIC) for a model with within-studies variance constrained to zero were lowest out of all models compared, there was no advantage to the multilevel approach. Therefore a random-effects meta-analysis, which only includes a between-studies variance component, was conducted.

**Table 3**. Comparing the fit of different multi-level models

|  | df | AIC | BIC | ll | LRT | p |
|---|---|---|---|---|---|---|
| Full three-level model | 3 | 79.86 | 84.83 | -36.92 |  |  |
| Between-studies variance constrained | 2 | 93.43 | 96.76 | -44.72 | 15.59 | 0.000 |
| Within-studies variance constrained | 2 | 77.84 | 81.17 | -36.92 | 0.00 | 1.000 |
| Both variance components constrained | 1 | 256.27 | 257.94 | -127.14 | 180.43 | 0.000 |

*Summary effect size*

The summary effect from random-effect meta-analysis was significantly different from zero, $\gamma = -0.30, p < 0.01$, 95% CI$[-0.50, -0.10]$. The random effect was also significant, indicating that there was residual heterogeneity between studies, $\tau^2 = 0.31, SE = 0.09, Q_{resid}(39) = 277.97, p < 0.01$. This is reflective of the diversity of studies included in the meta-analysis. See Figure 1 for a forest plot of the included studies.

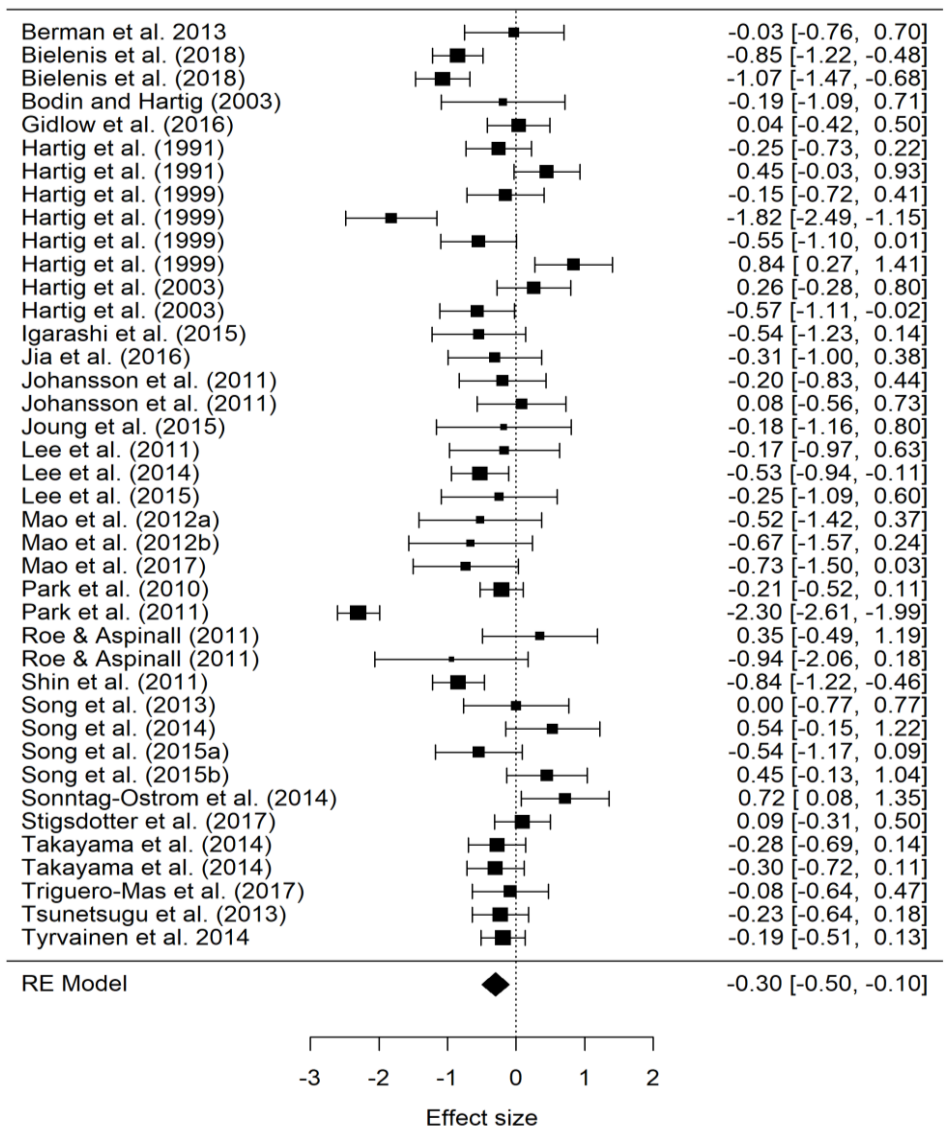

**Figure 1.** Forest plot of study effect sizes

*Publication bias*

Visual inspection of the funnel plot in Figure 2 was inconclusive with regard to publication bias. There was a lack of studies in both lower corners of the funnel plot, indicating that most small-sample studies reported effects with values close to the average weighted effect size. There were also some relatively high-powered studies with large, negative effects. However, Beggs' test of funnel asymmetry was non-significant ($z = 0.93, p = 0.35$). File drawer analysis indicated that 846 unpublished or unretrieved studies averaging null results would have to be added to render the average unweighted effect size non-significant. Thus, the extent of publication bias was hard to ascertain.

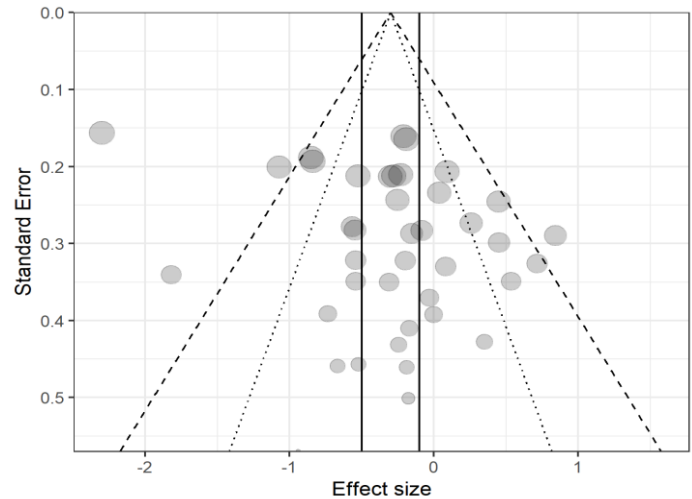

**Figure 2.** Funnel plot to assess potential publication bias

*Moderation analysis*

To investigate the source of heterogeneity, a random-effects MetaForest analysis was conducted with 10,000 iterations and replicated 100 times to ensure the reliability of findings. The replicated variable importance metrics can be seen in Figure S4. All variables that reduced the predictive performance of the model were dropped and the remaining eight carried forward to optimize the model. The estimated predictive performance in new data was positive; cross-validated $R^2_{cv} = 0.42$, out-of-bag $R^2_{oob} = 0.05$. The relative variable importance of the moderators in the final model is shown in Figure S5. The model identifies the proportion of females in the sample, the type of natural and built environment, the type of effect size, the time between natural and built environment visits, the country of study origin, the gender mix of the sample, and whether or not a baseline measurement was taken to be the most important moderators of the effect size from the 20 that were entered.

Partial dependency plots (Figure S6) were produced to examine the influence of each moderator on the effect size, while averaging over all other moderators. The model predicts that for a sample with a lower proportion of women, the effect size is larger. The effect size was also larger for agricultural, biodiverse and forest environments, compared to a park environment. Categories within other moderators showed similar relationships with the effect size. The eight most important moderators were entered into a meta-regression (Table 4), however, none were significant in a linear model.

**Table 4**. Meta regression model with most important moderators

| Variable | Estimate | SE | Z | p | CI |
| --- | --- | --- | --- | --- | --- |
| Intercept | -0.09 | 0.32 | -0.28 | 0.78 | [-0.72, 0.54] |
| Female | 0.01 | 0.01 | 0.60 | 0.55 | [-0.02, 0.04] |
| Natural environment: ABF vs Park | -0.14 | 0.20 | -0.70 | 0.48 | [-0.53, 0.25] |
| Natural environment: A vs BF | -0.35 | 0.49 | -0.71 | 0.48 | [-1.31, 0.61] |
| Natural environment: B vs F | 0.07 | 0.31 | 0.22 | 0.83 | [-0.54, 0.68] |
| Type of ES: SMCR vs SMD | 0.15 | 0.18 | 0.87 | 0.38 | [-0.19, 0.50] |
| Time between environments: Linear | 0.97 | 0.62 | 1.57 | 0.12 | [-0.24, 2.19] |
| Time between environments: Quad. | 0.58 | 0.50 | 1.17 | 0.24 | [-0.39, 1.56] |
| Time between environments: Cubic | -1.02 | 0.60 | -1.71 | 0.09 | [-2.19, 0.15] |
| Country: Asia vs USEurope | 0.14 | 0.16 | 0.90 | 0.37 | [-0.17, 0.45] |
| Country: US vs Europe | -0.08 | 0.27 | -0.29 | 0.77 | [-0.62, 0.46] |
| Sex: Mixed vs FemaleMale | -0.79 | 1.11 | -0.72 | 0.47 | [-2.96, 1.37] |
| Sex: Male vs Female | -2.16 | 2.11 | -1.03 | 0.30 | [-6.30, 1.97] |
| Baseline measurement: No vs Yes | 0.13 | 0.22 | 0.61 | 0.54 | [-0.29, 0.55] |
| Built environment: DowntownOther vs Residential | 0.20 | 0.27 | 0.72 | 0.47 | [-0.34, 0.73] |
| Built environment: Downtown vs Other | -0.40 | 0.51 | -0.78 | 0.43 | [-1.40, 0.60] |

Quality of evidence

A summary of findings table is presented in Figure S3. For randomised studies (*n*=25), initial confidence is high. However, the studies were downgraded due to serious risk of bias, inconsistency between studies, and plausible confounding. It was judged that because all studies had received a serious risk of bias in the individual study assessments, the overall body evidence would equally be deemed to have a serious risk of bias. Concerning the inconsistency domain, during the meta-analysis it was found that there was significant residual heterogeneity, therefore this was also marked with serious concerns. Lastly, a number of confounding variables were identified during the study-level bias assessments, and so this judgement was also applied to the body of evidence as a whole. In line with GRADE guidelines, the non-randomised studies ($n$=8) started as low quality due to residual confounding. In addition to the aforementioned judgements, these studies were further downgraded for imprecision due to small sample sizes and wide confidence intervals.

Publication bias, an overall large effect and a dose response gradient were not identified. Overall, the randomised studies were deemed to be of low quality and the non-randomised studies of very low quality.

## 4 Discussion

In this review and meta-analysis, 33 studies that investigated the effect of direct, short-term exposure to the natural environment on depressive mood were synthesised. Effect sizes ranged from -2.30 to 0.84, with an unweighted mean effect size of -0.29. However, risk of bias and quality assessments determined the current evidence is highly biased and of poor quality. Confidence in our conclusions is therefore limited, and the summary effect must be interpreted with caution. The meta-analysis also revealed significant residual heterogeneity between studies, which remains largely unexplained following moderator analysis.

Nevertheless, this review is in line with a previous meta-analyses of the effect of natural environment exposure on mood ($r$=-0.12, McMahan & Estes, 2015). The results also complement previous systematic reviews on the mental health benefits of, for example, forest bathing, horticultural activities, and green exercise (Bowler, Buyung-Ali, Knight, & Pullin, 2010; Hansen et al., 2017; Lee et al., 2017; Oh et al., 2017). In particular the findings converge with previous research into the presence of a dose-response relationship, which has demonstrated a boost to mood following a short period of exposure to nature. The most common exposure time identified in this review was 15 minutes - the experiments of Barton and Pretty 2010; and Shanahan et al. 2015, 2016) showed changes in mood following 5 minutes and 30 minutes of exposure time respectively. Overall, the review contributes to a growing evidence base concerning the mental health benefits of exposure to nature.

During moderator analysis using the MetaForest approach, it was found that proportion of female participants, type of natural environment and built environment, time between environments, baseline measurement, region of study origin, gender mix of the sample and type of effect size were important moderators of the effect size. This analysis draws some similarities with the results of a previous meta-analysis that found that type of emotion assessment, type of exposure to nature, location of study, and mean age significantly moderated the effect of nature on positive mood (McMahan & Estes, 2015). In addition the current analysis finds several between-study moderators to be relevant, which is reflective of the diversity of included studies. On the other hand, none were significant when entered in a meta-regression.

The lack of significance might be explained by the potential for bias in the included studies. A number of confounders were identified which may have influenced the results. It is not known to what extent carryover effects, whereby the effect of one environment might be 'carried over' to the next, might contribute to results: a wide range in the duration of time between environments was found. Another issue more generally relates to the issue of blinding participants and outcome assessors. It is essentially impossible to blind persons involved in interventions of this kind, since awareness of the environment is necessary. Van den Berg (2017) explains that these issues represent a key challenge in encouraging 'green prescriptions' (greening a person's environment, or taking them to a green environment, in order to promote health). Health professionals are inclined toward the results of randomised controlled trials, and often this approach is not appropriate for a nature-based intervention.

There were attempts to reduce bias in some studies. For example, two studies did not give prior warning of the order in which environments would be visited (Gidlow et al., 2016; Sonntag-Öström et al., 2014). These two studies also reported a clear process of randomisation. No study protocols could be found for studies included in this review, however a recent study of a park prescription program has done this (Razani et al., 2018). It is recommended that future research in the area take steps to reduce bias and improve quality where possible, in order to build a strong clinical evidence base. This will work to persuade policymakers and health professionals of the mental health benefits of exposure to nature.

Strengths and limitations

This review provides the most comprehensive and up-to-date findings on the effect of short-term exposure to the natural environment on depressive mood. Its key strengths are its broad range of included studies, and a fully reproducible and transparent meta-analysis.

However, this review also had some limitations. First, it was limited to English articles only. This prevented articles written in other languages from being included, however, a previous review that included relevant articles written in Korean did not find dissimilar results to this review (Lee et al., 2017). Second, it was not possible to retrieve data for two studies to enter into the meta-analysis. Both studies reported a reduction in depressive mood following exposure to the natural environment (Li et al., 2016; Perkins et al., 2011). Lastly, this review was concerned with short-term exposure only and does not address long-term effects of repeated exposure. It is assumed that repeated exposures would be cumulatively beneficial, and indeed a recent review found that long-term exposure to increased green and blue space in the residential environment is associated with improved mental health (Gascon et al., 2015).

Future research

Three suggestions for further research are made. First, future meta-analysis would benefit from improved descriptions and reporting of studies. For example, studies should provide an objective description of the experimental and control environments. This might be achieved by measuring the Normalized Difference Vegetation Index of the area, making use of street view imagery, or calculating percent tree canopy. Next, a detailed procedural description is required to fully understand the environmental context within which participants are placed. Moreover, it is recommended that appropriate guidelines are followed during reporting, for example, CONSORT (Schulz, Altman, & Moher, 2010) for randomised trials and TREND (Des Jarlais, Lyles, & Crepaz, 2004) for non-randomised studies. This ensures studies are fully described in a standardised manner.

Second, the MetaForest analysis revealed eight moderators that were associated with the effect size. In particular, the type of natural environment and proportion of females in the sample were the two most important moderators. The partial dependent plots showed that a larger effect size was associated with a lower proportion of women, and also in agricultural, biodiverse and forest environments, compared to the park environment. Further, the majority of studies also had young, usually male, university students as their participants. This reduces generalisability to other populations. It is therefore suggested that future research continues to explore the potential moderating role of type of environment and type of population group. This is important to understand in order to develop effective interventions to promote mood.

Lastly, increasing research is applying technology such as Global Positioning System, wearables, and ecological momentary assessment to investigate mental state over time and space (Bakolis et al., 2018; Birenboim, Dijst, Scheepers, Poelman, & Helbich, 2019; Chaix, 2018; Helbich, 2018). This represents the next step in this field of research whereby pre- and post-measures can be reformulated into a more dynamic approach. This removes the need for experimental procedure as participants can be followed in their daily life, and the effects of varying exposure duration and potential accumulation effects and long-term mental health benefits might be considered.

## 5 Conclusions

This review and meta-analysis finds a reduction in depressive mood following short-term exposure to the natural environment, however, studies were highly biased and of low quality. It is therefore unclear whether these findings would be replicated in higher quality studies. No significant moderators of the effect size were identified. More rigorous studies are required to improve our understanding of the relationship between the natural environment and mood.

**Acknowledgements**

We wish to thank all authors that provided additional data and/or clarification to complete the meta-analysis, namely: Christopher Gidlow, Terry Hartig, Gemma Hurst, Juyoung Lee, Genxiang Mao, Yoshifumi Miyazaki, Jenny Roe, Russ Searight, Chorong Song, Norimasa Takayama, Margarita Triguero-Mas.

**Funding**

This project has received funding from the European Research Council (ERC) under the European Union's Horizon 2020 research and innovation programme (grant agreement No. 714993).

**Declaration of interest**

None.

**Conflict of interest**

The authors have no conflict of interest to declare.

SUPPLEMENTARY MATERIALS

Table S1. PRISMA checklist

| Section/topic | # | Checklist item | Reported on page # |
|---|---|---|---|
| **TITLE** | | | |
| Title | 1 | Identify the report as a systematic review, meta-analysis, or both. | 1 |
| **ABSTRACT** | | | |
| Structured summary | 2 | Provide a structured summary including, as applicable: background; objectives; data sources; study eligibility criteria, participants, and interventions; study appraisal and synthesis methods; results; limitations; conclusions and implications of key findings; systematic review registration number. | 2 |
| **INTRODUCTION** | | | |
| Rationale | 3 | Describe the rationale for the review in the context of what is already known. | 3-4 |
| Objectives | 4 | Provide an explicit statement of questions being addressed with reference to participants, interventions, comparisons, outcomes, and study design (PICOS). | 5 |
| **METHODS** | | | |
| Protocol and registration | 5 | Indicate if a review protocol exists, if and where it can be accessed (e.g., Web address), and, if available, provide registration information including registration number. | 5 |
| Eligibility criteria | 6 | Specify study characteristics (e.g., PICOS, length of follow-up) and report characteristics (e.g., years considered, language, publication status) used as criteria for eligibility, giving rationale. | 5 |
| Information sources | 7 | Describe all information sources (e.g., databases with dates of coverage, contact with study authors to identify additional studies) in the search and date last searched. | 6 |
| Search | 8 | Present full electronic search strategy for at least one database, including any limits used, such that it could be repeated. | Supp material |
| Study selection | 9 | State the process for selecting studies (i.e., screening, eligibility, included in systematic review, and, if applicable, included in the meta-analysis). | 6 |
| Data collection process | 10 | Describe method of data extraction from reports (e.g., piloted forms, independently, in duplicate) and any processes for obtaining and confirming data from investigators. | 6 |

| Data items | 11 | List and define all variables for which data were sought (e.g., PICOS, funding sources) and any assumptions and simplifications made. | 6 |
|---|---|---|---|
| Risk of bias in individual studies | 12 | Describe methods used for assessing risk of bias of individual studies (including specification of whether this was done at the study or outcome level), and how this information is to be used in any data synthesis. | 6-7 |
| Summary measures | 13 | State the principal summary measures (e.g., risk ratio, difference in means). | 7 |
| **Section/topic** | **#** | **Checklist item** | **Reported on page #** |
| Synthesis of results | 14 | Describe the methods of handling data and combining results of studies, if done, including measures of consistency (e.g., I$^2$) for each meta-analysis. | 7-8 |
| Risk of bias across studies | 15 | Specify any assessment of risk of bias that may affect the cumulative evidence (e.g., publication bias, selective reporting within studies). | 6-7,9 |
| Additional analyses | 16 | Describe methods of additional analyses (e.g., sensitivity or subgroup analyses, meta-regression), if done, indicating which were pre-specified. | 7-8 |
| **RESULTS** | | | |
| Study selection | 17 | Give numbers of studies screened, assessed for eligibility, and included in the review, with reasons for exclusions at each stage, ideally with a flow diagram. | 10 + Supp material |
| Study characteristics | 18 | For each study, present characteristics for which data were extracted (e.g., study size, PICOS, follow-up period) and provide the citations. | 10-18 |
| Risk of bias within studies | 19 | Present data on risk of bias of each study and, if available, any outcome level assessment (see item 12). | 18-19 |
| Results of individual studies | 20 | For all outcomes considered (benefits or harms), present, for each study: (a) simple summary data for each intervention group (b) effect estimates and confidence intervals, ideally with a forest plot. | 21-23 |
| Synthesis of results | 21 | Present results of each meta-analysis done, including confidence intervals and measures of consistency. | 21-23 |
| Risk of bias across studies | 22 | Present results of any assessment of risk of bias across studies (see Item 15). | 23+26 |
| Additional analysis | 23 | Give results of additional analyses, if done (e.g., sensitivity or subgroup analyses, meta-regression [see Item 16]). | 24-26+Supp Material |
| **DISCUSSION** | | | |
| Summary of evidence | 24 | Summarize the main findings including the strength of evidence for each main outcome; consider their relevance to key groups (e.g., healthcare providers, users, and policy makers). | 26-27 |

| Limitations | 25 | Discuss limitations at study and outcome level (e.g., risk of bias), and at review-level (e.g., incomplete retrieval of identified research, reporting bias). | 27-28 |
|---|---|---|---|
| Conclusions | 26 | Provide a general interpretation of the results in the context of other evidence, and implications for future research. | 26-29 |
| **FUNDING** | | | |
| Funding | 27 | Describe sources of funding for the systematic review and other support (e.g., supply of data); role of funders for the systematic review. | 29 |

**Table S2.** ROBINS-I assessments

| | Confounding | Participant selection | Classification of interventions | Deviation from intended intervention | Missing data | Measurement of outcomes | Reporting bias | Overall bias |
|---|---|---|---|---|---|---|---|---|
| Joung et al. 2015 | No information | Low | Low | Low | Low | Serious | Moderate | Serious |
| Li et al., (2016) | Serious | Low | Low | Low | Low | Serious | Serious | Serious |
| Roe & Aspinall (2011) | Moderate | Low | Low | Low | Serious | Serious | Moderate | Serious |
| Song et al., (2013) | Serious | Low | Low | Low | Low | Serious | Serious | Serious |
| Song et al., (2014) | Moderate | Low | Low | Low | Low | Serious | Moderate | Serious |
| Song et al., (2015b) | Moderate | Low | Low | Low | Low | Serious | Moderate | Serious |
| Stigsdotter et al., (2017) | Serious | Low | Low | Low | Low | Serious | Moderate | Serious |
| Tsunetsugu et al. (2013) | No information | Low | Low | Low | Low | Serious | Moderate | Serious |

**Figure S1.** Flow diagram of study selection

**PRISMA 2009 Flow Diagram**

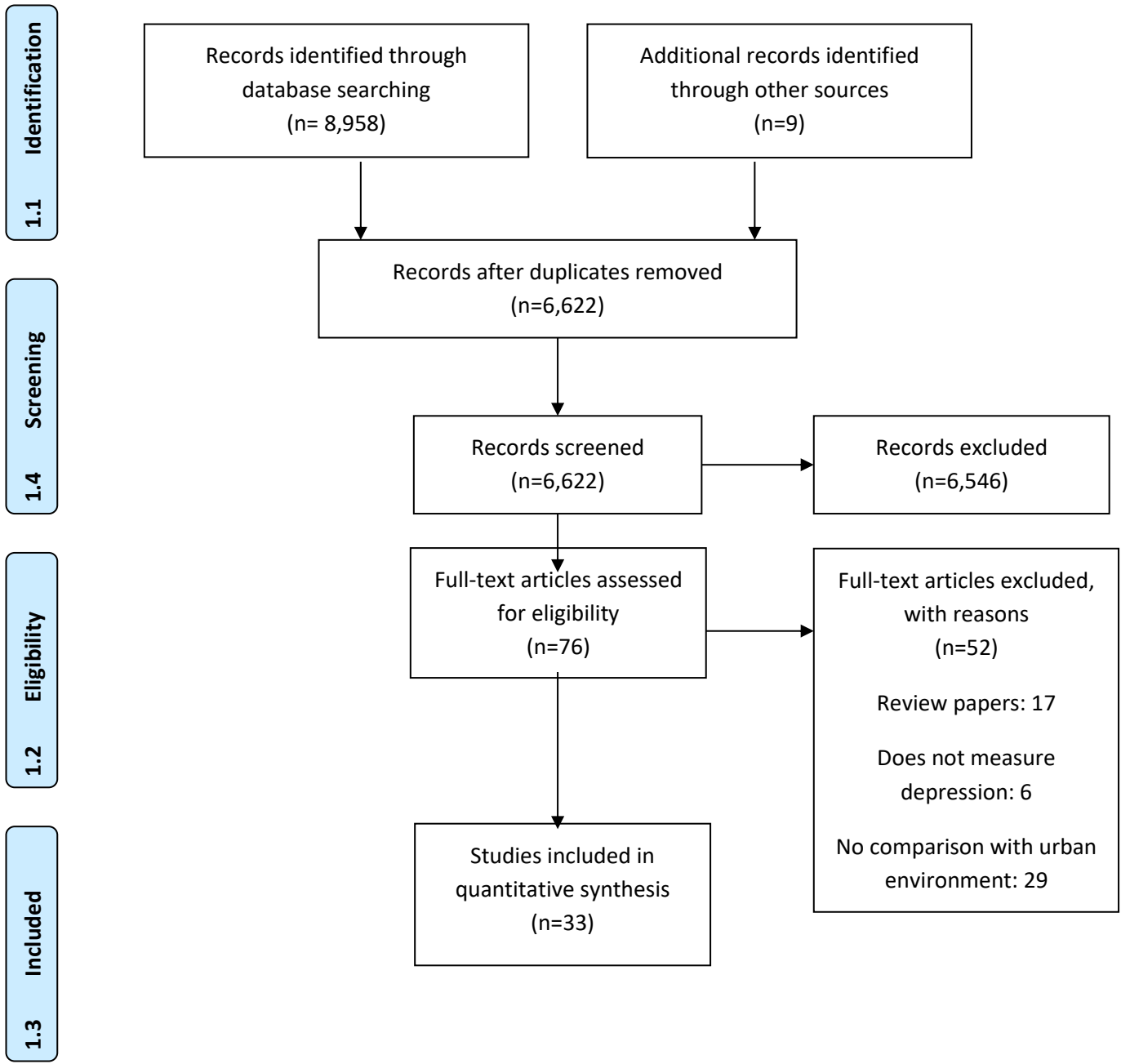

**Figure S2.** Risk of bias graph and table

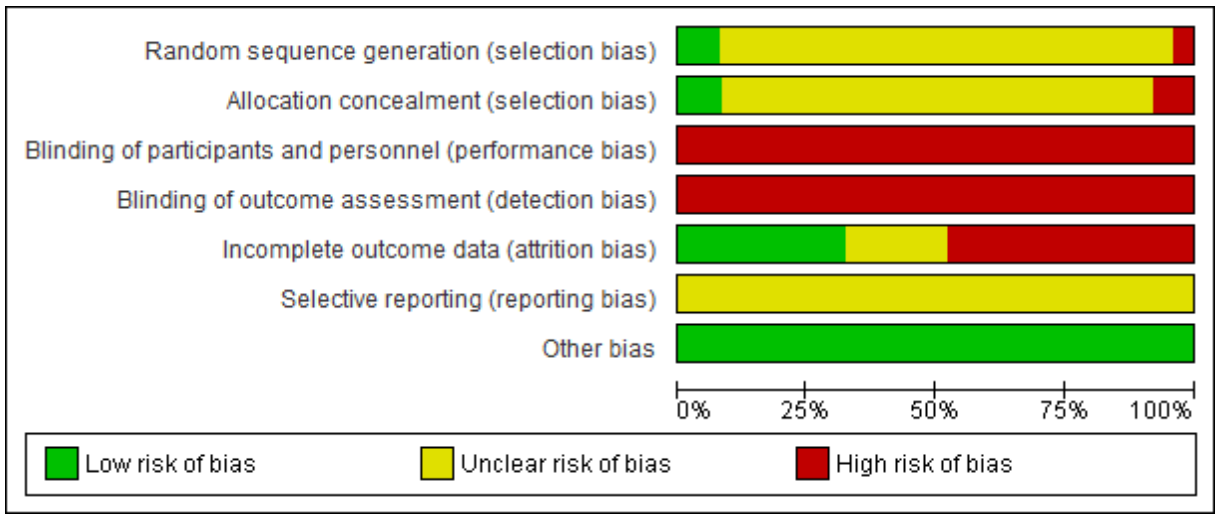

| Study | Random sequence generation (selection bias) | Allocation concealment (selection bias) | Blinding of participants and personnel (performance bias) | Blinding of outcome assessment (detection bias) | Incomplete outcome data (attrition bias) | Selective reporting (reporting bias) | Other bias |
|---|---|---|---|---|---|---|---|
| Berman et al. 2013 | ? | ? | − | − | − | ? | + |
| Bielinis 2018 | ? | − | − | − | + | ? | + |
| Bodin & Hartig 2003 | ? | ? | − | − | + | ? | + |
| Gidlow et al. 2016 | + | + | − | − | − | ? | + |
| Hartig et al. 1991 | ? | ? | − | − | − | ? | + |
| Hartig et al. 1999 | ? | ? | − | − | − | ? | + |
| Hartig et al. 2003 | ? | ? | − | − | − | ? | + |
| Igarashi et al. 2015 | ? | ? | − | − | + | ? | + |
| Jia et al 2016 | ? | ? | − | − | − | ? | + |
| Johansson et al. 2011 | ? | ? | − | − | + | ? | + |
| Lee et al. 2011 | ? | ? | − | − | + | ? | + |
| Lee et al. 2014 | ? | ? | − | − | − | ? | + |
| Lee et al. 2015 | ? | ? | − | − | − | ? | + |
| Mao et al. 2012a | ? | ? | − | − | + | ? | + |
| Mao et al. 2012b | ? | ? | − | − | ? | ? | + |
| Mao et al. 2017 | ? | ? | − | − | − | ? | + |
| Park et al. 2010 | ? | ? | − | − | − | ? | + |
| Park et al. 2011 | ? | ? | − | − | − | ? | + |
| Perkins et al. 2012 | ? | ? | − | − | ? | ? | + |
| Shin et al. 2011 | ? | ? | − | − | ? | ? | + |
| Song et al. 2015a | ? | ? | − | − | + | ? | + |
| Sonntag-Ostrom et al. 2014 | + | + | − | − | ? | ? | + |
| Takayama et al. 2014 | ? | ? | − | − | ? | ? | + |
| Triguero-Mas et al. 2017 | − | − | − | − | + | ? | + |
| Tyrvainen et al. 2014 | ? | ? | − | − | − | ? | + |

**Figure S3.** Summary of Findings table

| № of studies | Study design | Risk of bias | Inconsistency | Indirectness | Imprecision | Other considerations | the natural environment | the urban environment | Relative (95% CI) | Absolute (95% CI) | Certainty | Importance |
|---|---|---|---|---|---|---|---|---|---|---|---|---|
| **Certainty assessment** ||||||| **№ of patients** || **Effect** |||||

| № of studies | Study design | Risk of bias | Inconsistency | Indirectness | Imprecision | Other considerations | the natural environment | the urban environment | Relative (95% CI) | Absolute (95% CI) | Certainty | Importance |
|---|---|---|---|---|---|---|---|---|---|---|---|---|
| **Depressive mood** |||||||||||||
| 25 | randomised trials | very serious [a] | serious [b] | not serious | not serious | all plausible residual confounding would suggest spurious effect, while no effect was observed | 912 | 902 | - | SMD **0.05 SD higher** (0.4 lower to 0.5 higher) | ⊕⊕◯◯ LOW | IMPORTANT |
| **Depressive mood** |||||||||||||
| 8 | observational studies | very serious [a] | serious [b] | not serious | serious [c] | all plausible residual confounding would suggest spurious effect, while no effect was observed | 179 | 175 | - | SMD **0.38 SD lower** (0.59 lower to 0.16 lower) | ⊕◯◯◯ VERY LOW | IMPORTANT |

**CI:** Confidence interval; **SMD:** Standardised mean difference

*Explanations*

a. Issues with randomisation method, allocation concealment, blinding, incomplete outcome data and selective reporting.

b. Random effects meta-analysis found residual heterogeneity between studies.

c. Small sample size and wide confidence intervals.

**Figure S4.** Replicated MetaForest for variable preselection

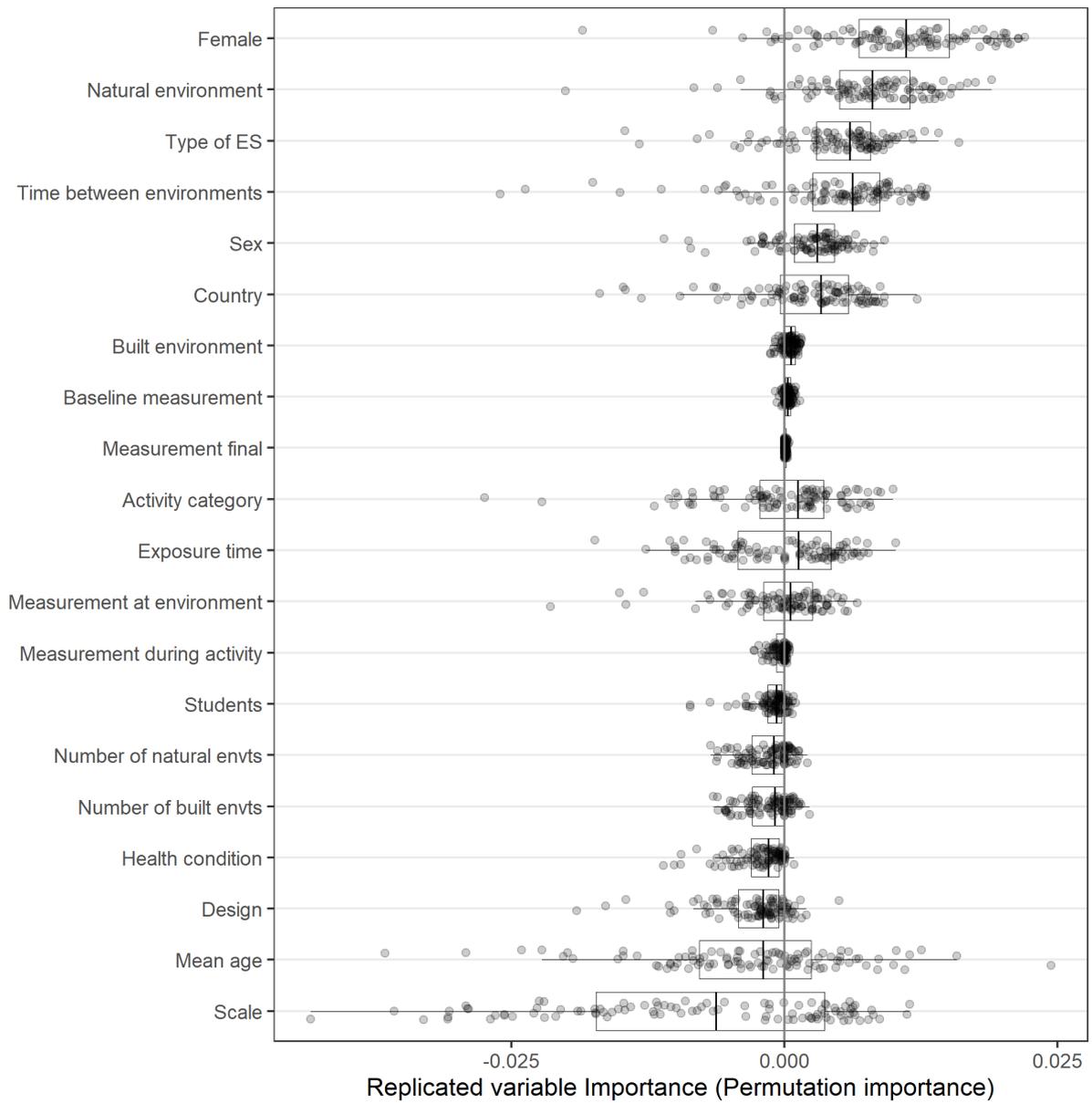

**Figure S5.** Variable importance for final model

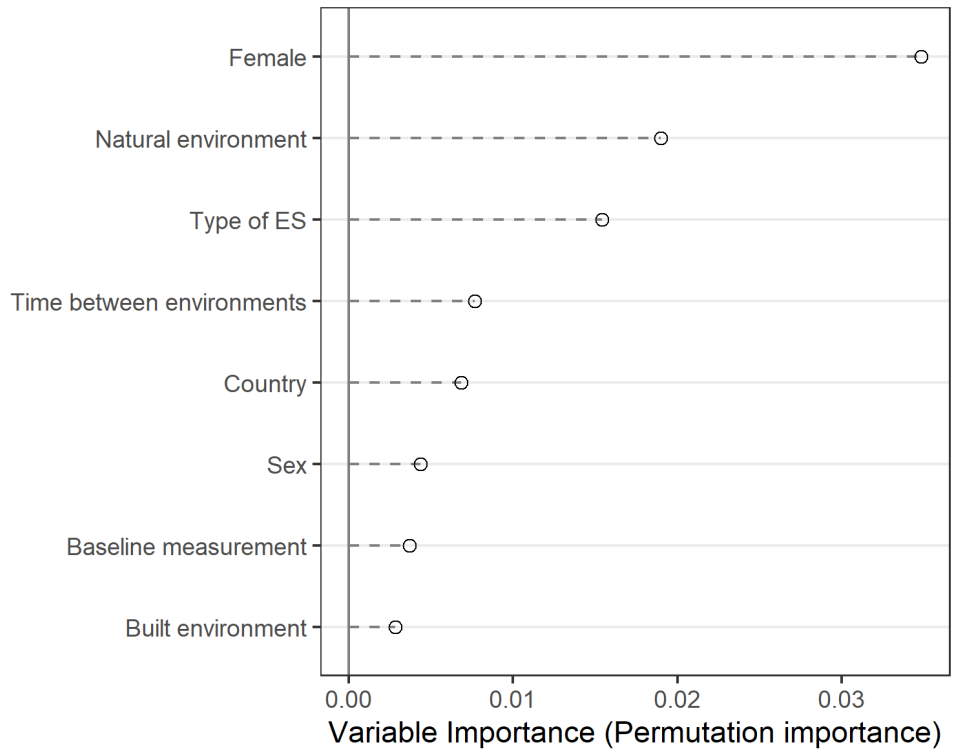

*Note.* The MetaForest analysis was tuned using the eight moderators above. The optimal model used uniform weights, 2 candidate variables at each split, and a minimal terminal node size of 3.

**Figure S6.** Partial dependence plots

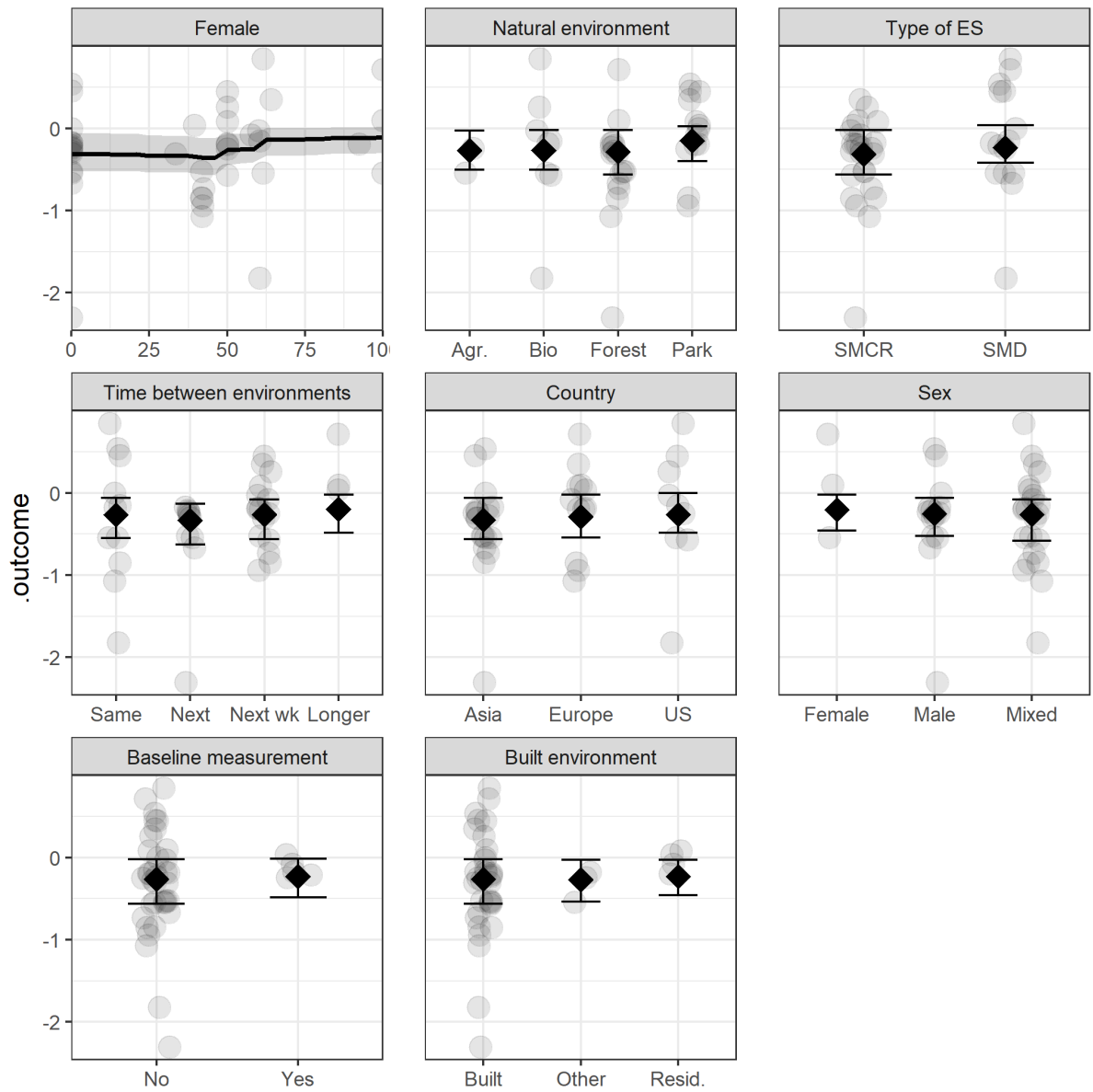

# Search strategies for all databases searched

## MEDLINE

(((((people[MeSH Major Topic] OR "people" OR adults[MeSH Major Topic] OR "adults" OR volunteers[MeSH Major Topic] OR "volunteers" OR participants OR subjects OR students OR respondents))

AND

(intervention[MeSH Major Topic] OR "intervention" OR experiment OR randomized[MeSH Major Topic] OR "randomized" OR crossover OR "case-crossover" OR "pre-post" OR comparison OR "non-randomized" OR "quasi-experiment" OR controlled OR control OR "control group"))

AND

("natural environment" OR "natural outdoor environment" OR outdoors OR outside OR "green space" OR forest[MeSH Major Topic] OR "forest" OR "woodland" OR garden OR allotment OR countryside OR "open space" OR landscape OR parkland OR park NOT parkin*))

AND

("mental health" OR depression[MeSH Major Topic] OR "depression" OR mood[MeSH Major Topic] OR "mood" OR "mood disorder" OR "mood change" OR "major depressive disorder" OR "depressive symptoms")

Sort by: Author Filters: Humans; English

## PsychINFO via OvidSP

people.mp.
adults.mp.
volunteers.mp.
experimental subjects/
participants.mp.
subjects.mp.
students.mp.
respondents.mp. [mp=title, abstract, heading word, table of contents, key concepts, original title, tests & measures]
1 or 2 or 3 or 4 or 5 or 6 or 7 or 8
intervention/
intervention.mp.
experiment.mp. [mp=title, abstract, heading word, table of contents, key concepts, original title, tests & measures]
randomized.mp.
crossover.mp.

"case-crossover".mp.

"pre-post".mp.

comparison.mp. [mp=title, abstract, heading word, table of contents, key concepts, original title, tests & measures]

"non-randomized".mp.

controlled.mp. [mp=title, abstract, heading word, table of contents, key concepts, original title, tests & measures]

control.mp. [mp=title, abstract, heading word, table of contents, key concepts, original title, tests & measures]

"control group".mp. [mp=title, abstract, heading word, table of contents, key concepts, original title, tests & measures]

10 or 11 or 12 or 13 or 14 or 15 or 16 or 17 or 18 or 19 or 20 or 21

forest.mp.

woodland.mp.

"natural environment".mp.

"natural outdoor environment".mp.

outdoors.mp. [mp=title, abstract, heading word, table of contents, key concepts, original title, tests & measures]

outside.mp. [mp=title, abstract, heading word, table of contents, key concepts, original title, tests & measures]

"green space".mp.

garden.mp.

allotment.mp.

countryside.mp. [mp=title, abstract, heading word, table of contents, key concepts, original title, tests & measures]

(park not parkin*).mp. [mp=title, abstract, heading word, table of contents, key concepts, original title, tests & measures]

parkland.mp. [mp=title, abstract, heading word, table of contents, key concepts, original title, tests & measures]

"open space".mp.

landscape.mp. [mp=title, abstract, heading word, table of contents, key concepts, original title, tests & measures]

"urban area".mp. [mp=title, abstract, heading word, table of contents, key concepts, original title, tests & measures]

"urban environment".mp. [mp=title, abstract, heading word, table of contents, key concepts, original title, tests & measures]

urban environments/

exp Built Environment/

23 or 24 or 25 or 26 or 27 or 28 or 29 or 30 or 31 or 32 or 33 or 34 or 35 or 36 or 37 or 38 or 39 or 40

"mental health".mp. [mp=title, abstract, heading word, table of contents, key concepts, original title, tests & measures]

"depression (emotion)"/

exp MAJOR DEPRESSION/

depression.mp.

mood.mp.

"mood change".mp. [mp=title, abstract, heading word, table of contents, key concepts, original title, tests & measures]

"depressive symptoms".mp. [mp=title, abstract, heading word, table of contents, key concepts, original title, tests & measures]

"mood disorder".mp. [mp=title, abstract, heading word, table of contents, key concepts, original title, tests & measures]

42 or 43 or 44 or 45 or 46 or 47 or 48 or 49

9 and 22 and 41 and 50

**EMBASE**

#1    participants OR 'students'/exp OR students OR subjects OR 'volunteer'/exp OR 'volunteer' OR people OR respondents

#2    'control group' OR control OR 'non-randomized' OR 'controlled clinical trial (topic)' OR 'pre-post' OR comparison OR 'case-crossover' OR 'crossover procedure' OR 'randomized' OR 'randomized controlled trial' OR 'experiment' OR 'intervention study'

#3    ('open space' OR 'allotment' OR 'garden' OR 'green space' OR 'natural outdoor environment' OR 'natural environment' OR outdoors OR outside OR 'forest' OR woodland OR countryside OR landscape OR 'recreational park' OR parkland OR 'urban area' OR 'urban environment') NOT parkin*

#4    'mental health' OR 'mood change' OR 'depressive symptoms' OR 'mood' OR 'mood disorder' OR 'major depression' OR 'depression'

#1 AND #2 AND #3 AND #4 AND [humans]/lim AND [english]/lim

**Scopus**

TITLE-ABS-KEY ( people OR volunteers OR subjects OR students OR participants OR respondents ) AND ( TITLE-ABS-KEY ( intervention OR experiment OR "randomized controlled trial" OR randomized OR crossover OR "case-crossover" OR pre-post OR comparison OR non-randomized OR exposure OR controlled OR control OR "control group" ) ) AND ( TITLE-ABS-KEY ( "green space" OR "natural outdoor environment" OR "natural environment" OR outdoors OR outside OR "open space" OR countryside OR allotment OR garden OR forest OR woodland OR landscape OR parkland OR park AND NOT parkin* ) ) AND ( TITLE-ABS-KEY ( "mental health" OR depression OR depressive OR "depressive symptoms" OR mood OR "mood disorder" ) ) AND ( LIMIT-TO ( LANGUAGE , "English " ) )

**Web of Science**

(participants OR students OR subjects OR volunteers OR people) AND

('control group' OR 'non-randomized' OR 'pre-post' OR 'case-crossover' OR 'crossover procedure' OR randomized OR 'randomized controlled trial' OR intervention) AND

(park OR 'open space' OR allotment OR garden OR 'green space' OR 'natural outdoor environment' OR 'natural environment' OR forest OR woodland) AND

('mood change' OR 'depressive symptoms' OR mood OR 'mood disorder' OR 'major depression' OR depression)